\documentclass[12pt]{iopart}

\usepackage{bm}
\usepackage{epsfig}
\usepackage{hyperref}
\usepackage{graphicx}
\eqnobysec

\begin{document}


\begin{flushright}{\large \tt LAPTH-1236/08}
\end{flushright}

\title{How to constrain inflationary parameter space with minimal priors}

\author{Jan~Hamann, Julien Lesgourgues and
Wessel Valkenburg}

\address{LAPTH (Laboratoire d'Annecy-le-Vieux de Physique
Th\'eorique, CNRS UMR5108 \&~Universit\'e de Savoie), BP 110, F-74941
 Annecy-le-Vieux Cedex, France}

\ead{\mailto{jan.hamann@lapp.in2p3.fr},
     \mailto{julien.lesgourgues@lapp.in2p3.fr},
     \mailto{wessel.valkenburg@lapp.in2p3.fr}}

\begin{abstract}

We update constraints on the Hubble function $H(\phi)$ during
inflation, using the most recent cosmic microwave background (CMB)
and large scale structure (LSS) data. Our main focus is on a
comparison between various commonly used methods of calculating the
primordial power spectrum via analytical approximations and the
results obtained by integrating the exact equations numerically. In
each case, we impose na\"{i}ve, minimally restrictive priors on the
duration of inflation. We find that the choice of priors has an
impact on the results: the bounds on inflationary parameters can vary
by up to a factor two.  Nevertheless, it should be noted that within
the region allowed by the minimal prior of the exact method, the
accuracy of the approximations is sufficient for current data.  We
caution however that a careless minimal implementation of the
approximative methods allows models for which the assumptions behind
the analytical approximations fail, and recommend using the exact
numerical method for a self-consistent analysis of cosmological data.

\end{abstract}
\maketitle

\section{Introduction}\label{sec:introduction}

Cosmic inflation is the simplest and most robust paradigm capable of
providing self-consistent initial conditions to the Hot Big Bang
scenario
\cite{Starobinsky:1980te,Guth:1980zm,Sato:1980yn,Linde:1981mu,Albrecht:1982wi,Linde:1983gd},
as well as a mechanism for the quantum-gravitational generation of
primordial scalar (density) perturbations and gravitational waves
\cite{Starobinsky:1979ty,Mukhanov:1981xt,Hawking:1982cz,Starobinsky:1982ee,Guth:1982ec,Bardeen:1983qw,Abbott:1984fp}.
The Fourier power spectrum ${\cal P}_{\cal R}(k)$ of the former
is observed today in the cosmic microwave background (CMB) and the
large scale structure (LSS). Vice versa, at present the CMB and the
LSS provide the only quantifiable observables which can confirm or
falsify inflationary predictions. That is why matching concrete
inflationary models to observations has become one of the leading
quests in cosmology.

In the standard inflationary picture, the amplitude of perturbations
for a given comoving Fourier mode $k$ depends crucially on the
dynamics of inflation around the time of Hubble exit for this mode.
Each Hubble exit time is conveniently parameterised in terms of the
number of $e$-folds $N$ before inflation ends. The relation between $k$
and $N$ depends very much on the overall energy scale of inflation.
The ensemble of modes observable in the CMB and in the quasi-linear
part of the LSS power spectra corresponds to a range $\Delta N \sim 10$
called ``observable inflation''. The total duration of inflation is a
priori unlimited, but the number of $e$-folds between the time at which
the presently observable Universe became as large as the Hubble radius
and the end of inflation can only vary in the approximate range $30 <
N <60$, that will be called ``relevant inflation'' throughout this
paper.

The literature on inflation constraints is plethoric. For simplicity,
a majority of papers are restricted to the case in which inflation is
driven by a single scalar field $\phi$ (an inflaton) with a canonical
kinetic term, some potential $V(\phi)$ and minimal coupling to
Einstein gravity (however, each of these assumptions can be relaxed
and has already been studied separately). Traditional works are based
on the definition of spectral parameters (amplitude, index, possibly
running) for density perturbations and gravitational waves. In a first
step, these parameters are fitted to the data; in a second step, one
tries to infer the class of inflationary models compatible with
derived bounds on spectral parameters.

In the last years, many works have gone beyond this approach,
recognising that the introduction of spectral parameters puts already
a strong theoretical prior on the models, and is by no means a
necessary step. It is more realistic and equally efficient to fit
directly to the data the (more fundamental) parameters governing the
dynamics of inflation and/or the inflaton potential. Within the main
stream (standard single field inflation), recently published analyses
fall in two categories which are both interesting and complementary:
either one assumes a particular model based on a definite form for the
inflaton potential throughout relevant inflation, and derives
constraints on the free parameters of this potential (top-down
approach); or one employs a generic parameterisation of the potential
$V(\phi)$ or another function governing inflationary dynamics, e.g.
$H(\phi)$, and tries to reconstruct this function from the data
(bottom-up approach).  The second approach aims at avoiding
theoretical priors as much as possible, and concentrating on what the
data exactly tells us, although no parameterisation can be completely
general: sharp features are usually excluded {\it ab initio}
(obviously, all possible features cannot be accounted for with a
reasonable number of free parameters). Even within the bottom-up
approach, a distinction can be established between conservative
analyses reconstructing only the part of $V(\phi)$ corresponding to
observable inflation; and more aggressive analyses in which the
potential (or the function $H(\phi)$) is extrapolated till the end of
inflation, and subject to a prior on the minimum duration of relevant
inflation (e.g. $N \geq 30$). This more aggressive method should imply
varying many more parameters, since in this case the parameterisation
should be accurate over 30 to 60 $e$-folds instead of just $\sim \! 
10$. The fact of extrapolating is by itself an extra theoretical
prior, since cosmological data tell us essentially nothing about the
era between observable inflation and Nucleosynthesis: the end of
inflation could be subject to multi-field dynamics, experience phase
transitions, be split into several non-contiguous short inflationary
stages, etc.

Here we address only the most conservative approach, i.e., the
reconstruction of the inflationary dynamics during observable
inflation, with a minimal number of assumptions. After the publication
of WMAP results, this approach was followed in
Refs.~\cite{Peiris:2006ug,Easther:2006tv,Peiris:2006sj,Lesgourgues:2007gp,Lesgourgues:2007aa}.
It was stressed in~\cite{Peiris:2006ug,Easther:2006tv,Peiris:2006sj}
and~\cite{Lesgourgues:2007aa} that the quantity primarily constrained
by the data is $H(\phi)$: hence, this function is the one which should
be parameterised in some way and fitted to the data. The knowledge of
$H(\phi)$ uniquely defines the potential $V(\phi)$, and
Ref.~\cite{Lesgourgues:2007aa} presented the collection of potentials
$V(\phi)$ corresponding to the ensemble of functions $H(\phi)$ allowed
by current WMAP and SDSS LRG data. The results are still plagued by
degeneracies: since the energy scale of inflation is unknown, the data
favors a parametric family of inflaton potentials rather than a
precise shape within the observable window. However, a lot of
improvement is expected from the next generation of CMB experiments,
especially the {\sc planck} satellite. The reconstruction of
observable inflation will improve spectacularly if primordial
gravitational waves are observed by {\sc planck} or another
experiment, in the form of polarised $B$-modes. This would fix the
tensor-over-scalar ratio $r$, and hence the energy scale of inflation.
Even without $B$-modes, the Planck data would provide $r$-dependent
constraints on the inflation potential of unprecedented precision. In
this perspective, it is worth comparing the details and merits of each
reconstruction method.

Choosing to concentrate on the reconstruction of $H(\phi)$ during
observable inflation does not fix the method entirely, in particular
as far as the computation of the primordial spectra is concerned. The
authors of~\cite{Peiris:2006ug,Easther:2006tv,Peiris:2006sj} employed
analytic approximations of two different forms, while those
of~\cite{Lesgourgues:2007aa}\footnote{Numerical spectrum computations
were also employed in various complementary approaches to the problem
of constraining inflation, based on definite
potentials~\cite{Martin:2006rs,Covi:2006ci,Ringeval:2007am,Lorenz:2007ze}
or on a frequentist analysis with extrapolation of the potential
throughout relevant inflation~\cite{Powell:2007gu}.} wrote a module
appended to \texttt{CAMB}~\cite{Lewis:1999bs} and
\texttt{CosmoMC}~\cite{Lewis:2002ah}, which derives the numerical
spectra for each new set of inflationary parameters by numerically
solving the exact equations.  It is interesting to study whether the
difference between these methods is relevant given the precision of
current and future data. Beyond the issue of perturbations, different
methods could also differ through different assumptions concerning the
parameterisation of the background evolution, and the exact number of
$e$-folds during which this parameterisation is (explicitly or
implicitly) assumed to hold and be compatible with accelerated
expansion. The goal of this paper is to compare in details these
different techniques, and to see how each difference impacts the
constraints obtained from current data.  Although current inflaton
potential reconstructions are still dominated by degeneracies, a
careful understanding will be necessary before applying these methods
to the highly precise data expected in the next years.

In the next section we will briefly review the theory of inflationary
perturbations and discuss the exact approach, as well as two commonly
used approximative methods for the calculation of the primordial
perturbation spectra. In section \ref{sec:constraints}, we will
present the results of an analysis of current data and demonstrate
that the bounds using the approximative methods with a na\"ive prior
differ significantly from the constraints inferred with the exact
method. We will track down the cause of these differences and compare
the accuracy of the approximations in section \ref{sec:diff} before we
conclude in section
\ref{sec:conclusions}.

\section{Background and perturbations in single field inflation
\label{sec:perturbations}}

The observable spectra of density perturbations and gravitational
waves are directly related to the evolution of the Hubble parameter $H
\equiv \dot{a}/a$ as a function of $\phi$ in the neighbourhood of an
arbitrary pivot value $\phi_*$. The function $H(\phi-\phi_*)$ can in
principle be reconstructed from the data without any need to assume
an explicit value of $\phi_*$.
Each $H(\phi-\phi_*)$ defines a unique set 
$\left\{V(\phi-\phi_*),\dot \phi_{\rm ini}\right\}$ through
\begin{eqnarray}
  -\frac{32\pi^2}{M_{\rm
  Pl}^4}V(\phi-\phi_*)&=\left[H'(\phi-\phi_*)\right]^2-\frac{12\pi}{M_{\rm
  Pl}^2}H^2(\phi-\phi_*),\\ \quad \quad \quad \quad \quad \quad \quad
\dot{\phi} &= - \frac{M_{\rm Pl}^2}{4 \pi} \, H'(\phi-\phi_*), \label{eq:background}
\end{eqnarray}
whenever $\dot\phi\not= 0$~\footnote{Such a singularity is never
  reached as long as $H(\phi)$ is used as the defining quantity and
  has an analytic expression over the range considered. As mentioned
  in Ref.~\cite{Lesgourgues:2007aa}, $\dot\phi = 0$ can be reached for
  a field value $\phi_1$ only if $H$ has a non-analytical expression
  like $(H-H_1) \propto (\phi-\phi_1)^{3/2}$ in the vicinity of
  $\phi_1$. This cannot happen with the parametrisations used in this
  work (polynomial expressions for either $H(\phi)$ or
  $H^2(\phi)$). In addition, reaching $\dot\phi = 0$ would imply that
  the field changes direction with an opposite sign for $H'$, which is
  not compatible with the assumption of a single-valued function
  $H(\phi)$.}  (the prime denotes a derivative with respect to $\phi$,
and we have set $GM_{\rm Pl}^2=\hbar=c=1$). In
Ref.~\cite{Lesgourgues:2007aa}, the defining quantity $H(\phi-\phi_*)$
was Taylor-expanded up to the cubic term:
\begin{equation}
  H(\phi-\phi_*) = H_* + H_*'( \phi-\phi_*) + \frac{1}{2} H_*''( \phi-\phi_*)^2 +
  \frac{1}{6} H_*'''( \phi-\phi_*)^3~,
  \label{taylor}
\end{equation}
which is equivalent to keeping the first three slow-roll parameters
\begin{eqnarray}
  \epsilon & = \frac{M_{\rm Pl}^2}{4 \pi} \left[
    \frac{H'}{H} \right]^2,\\
  \eta & = \frac{M_{\rm Pl}^2}{4 \pi} \frac{H''}{H},\\
  \xi & = \frac{M_{\rm Pl}^4}{16 \pi ^2} \; \frac{H' H'''}{H^2},
\end{eqnarray}
in the Hubble flow hierarchy~\cite{Liddle:1994dx,Kinney:2002qn}, as in
the slow-roll reconstruction approach of
Refs.~\cite{Easther:2002rw,Peiris:2006ug,Easther:2006tv,Peiris:2006sj}.
Note that the universe expansion remains accelerated as long as
$\epsilon<1$.  For practical purposes, any parameterisation of
$H(\phi)$ could be used when fitting the data with a Bayesian MCMC
analysis. Besides the issue of priors on inflationary parameters, each
parameterisation corresponds to a different ensemble of possible
inflationary models. One can wonder how much the final results (i.e., the range
of allowed potentials) depends on the parameterisation. In the
following analysis we will compare the results obtained by
Taylor-expanding either $H(\phi)$ or $H^2(\phi)$ at the same order,
using in both cases the same flat priors on the first slow-roll
parameters expressed at the pivot scale $\phi_*$.

Once the ensemble of possible inflationary models has been specified,
the analysis still depends on the way to calculate the perturbation
spectra in single-field inflation, and on a theoretical prior on the
duration of inflation. In this section we will give a very brief
summary of three different approaches used
in~Refs.~\cite{Peiris:2006ug,Easther:2006tv,Peiris:2006sj,Lesgourgues:2007aa}.
Keep in mind that throughout this paper, we are working under the
assumption of a minimal prior. In other words, we do not impose any
lower bounds on the number of $e$-folds of inflation, we only demand
that the spectra of a model can be calculated with the respective
methods.  So, besides possible differences in the accuracy of the
resulting spectra, the default implementation of the three methods
will also differ in the range of parameter values that would be
excluded straight away.

\subsection{Exact spectra via mode equation}

Given $H$ as a function of $\phi$ during inflation, the spectrum of
curvature perturbations ${\cal P}_{\cal R}$ and gravitational waves
${\cal P}_{T}$ can be calculated exactly by integrating the
scalar/tensor {\it mode equation} (see, e.g,. \cite{Mukhanov:1990me}):
  \begin{equation}
    \frac{d^2 \xi_{\rm S,T}}{d\eta^2} + \left[ k^2 - \frac{1}{z_{\rm S,T}}
    \frac{d^2 z_{\rm S,T}}{d\eta^2}\right] \xi_{\rm S,T}=0
    \label{eq:mode}
  \end{equation}
with $\eta=\int dt/a(t)$ and $z_S= a\dot\phi/H$ for scalars, 
  $z_T= a$ for tensors. The evolution of
the background is determined by
\begin{equation}
  \dot{\phi} = - \frac{M_{\rm Pl}^2}{4 \pi} \, \frac{{\rm d} H}{{\rm
      d}\phi}.
\end{equation}
The $\xi_{\rm S,T}$ are usually taken to be in the Bunch-Davies vacuum
when they are well within the horizon, and their evolution needs to be
tracked until $|\xi_{\rm S,T}|/z_{\rm S,T}$ converges to a constant
value, in order to define the observable spectra:
\begin{equation}
  \frac{k^3}{2 \pi^2} \frac{|\xi_S|^2}{z_S^2} \rightarrow {\cal
  P}_{\cal R}~, \qquad \frac{32 k^3}{\pi M_{\rm Pl}^2}
  \frac{|\xi_T|^2}{z_T^2} \rightarrow {\cal P}_{T}~.
\end{equation}
In principle, observable inflation could be interrupted for a very
short amount of time, resulting in characteristic features in the
spectra. In the mainstream approach, this situation is not considered
for simplicity.  Actually, the numerical module used in
Ref.~\cite{Lesgourgues:2007aa} eliminates models violating $\epsilon
\leq 1$ at any point during the period of time the mode equation is
integrated. More precisely, for any wavenumber in the range $[k_{\rm
min}, k_{\rm max}]=[3\times 10^{-6},1.2]~$Mpc$^{-1}$ needed by
\texttt{CAMB} (the pivot scale being fixed at $k_*=0.01~$Mpc$^{-1}$),
the module integrates Eq.~(\ref{eq:mode}) from the time at which
$k/aH=50$ and until $[ {\rm d} \ln {\cal P}_{{\cal R}, T} / {\rm d}
\ln a ] < 10^{-3}$. If for a given function $H(\phi-\phi_*)$ the
product $aH$ does not grow monotonically by a sufficient amount for
fulfilling the above conditions, the model is rejected (we recall that
it is equivalent to impose that $aH$ grows or that $\epsilon$ is
greater than one).

The condition $[ {\rm d} \ln {\cal P}_{{\cal R}, T} / {\rm d} \ln a ]
< 10^{-3}$ is motivated by our desire to obtain a 0.1\% accuracy in
the power spectra.  The error made on ${\cal P}_{{\cal R}, T}$ by
stopping the integration of perturbations at a finite time can be
estimated analytically, comparing the amplitude of the decaying mode
to that of the non-decaying mode for ${\cal R}$ or gravitational waves during
inflation. The decaying over non-decaying mode ratio evolves in a
first approximation like $a^{-1}$, i.e., like $e^{-N}$. Hence, a few
lines of algebra show that the derivative $[ {\rm d} \ln {\cal P} /
{\rm d} \ln a ]$ is a good approximation for the relative error $[
\Delta {\cal P} / {\cal P} ]$ produced by stopping integration at a
finite time. Other parameters governing the precision of the power
spectra calculation (like the step of integration, the choice of the
initial integration time for each mode, etc.) where chosen in such way
that the above source of error is the dominant one.

The numerical evaluation of the spectrum involves solving equations
(\ref{eq:mode}) for each value of $k$, but this does not increase the
total running time of a Boltzmann code like \texttt{CAMB} by a noticeable
amount.  Nevertheless, there exist a number of approximations in the
literature, which simplify the calculation considerably.

\subsection{\label{subsec:apprI}Approximation~{\it I}}

This method was employed in \cite{Peiris:2006ug,Peiris:2006sj}, and
relies on the validity of the analytical slow-roll approximations,
\begin{eqnarray}
	\mathcal{P_R}(k) &\simeq \frac{\left[ 1 - 2(C_1 + 1) 
\epsilon + C_1 \eta \right]^2}{\pi \epsilon} \left. \left( \frac{H}{M_{\rm 
Pl}} \right)^2 \right|_{k=aH}, \label{eq:ps}\\
	\mathcal{P}_{\rm T}(k) &\simeq \left[ 1 - (C_1 + 1) \epsilon 
\right]^2 \; \frac{16}{\pi} \left. \left( \frac{H}{M_{\rm Pl}} 
\right)^2 \right|_{k=aH}, \label{eq:pt}
\end{eqnarray}
with $C_1 = -2 + \ln 2 + \gamma$, where $\gamma$ is the Euler-Mascheroni 
constant. These equations were first derived in \cite{Stewart:1993bc}
and are accurate only to first order in the slow-roll parameters,
assuming additionally that $\epsilon$ and $\eta$ are constant.
Here, one only needs to solve one differential equation to determine
$\phi(k)$,
\begin{equation}
	\label{eq:appi}
	\frac{{\rm d}\phi}{{\rm d}\ln k} = - \frac{M_{\rm Pl}}{2 \sqrt{ 
\pi}} \; \frac{\sqrt{\epsilon}}{1-\epsilon}, \label{eq:phiofk}
\end{equation}
assuming $\phi(k_*) = \phi_*$. Once $\phi(k)$ is known, the
slow-roll parameters and hence the spectrum can be evaluated for each
value $k$.
In this approach, the evolution of Eq.~(\ref{eq:phiofk}) has to be
followed throughout the observable range of wavelengths. If $\epsilon
\geq 1$, equation (\ref{eq:appi}) will diverge, so models with
$\epsilon > 1$ within this range will have to be excluded when using
this method. If however, the inflationary condition were violated just
before or after this range, the model would not be ruled out, and the
resulting spectra would likely be inaccurate.

\subsection{\label{subsec:apprII}Approximation~{\it II}}

This method is based on the usual Taylor-expansion of the spectra in
log-space around a pivot scale $k_*$ (see
e.g.,~\cite{Leach:2002ar,Leach:2003us}),
\begin{eqnarray}
  \ln \mathcal{P_R} &\simeq \ln A_{\rm S} + (n_{\rm S} - 1) \ln
  \left(k/k_*\right) + \frac{1}{2} \alpha_{\rm S}
  \left(\ln\left(k/k_*\right)\right)^2,\\
  \ln \mathcal{P}_{\rm T} &\simeq \ln  A_{\rm T} + n_{\rm T} \ln
  \left(k/k_*\right),
\end{eqnarray}
with the spectral indexes $n_{\rm S/T}$, and the running of the scalar
tilt $\alpha_{\rm S}$ given by their second-order slow-roll expressions
\begin{eqnarray}
  n_{\rm S} & \simeq 1 + 2 \eta - 4 \epsilon - 2 (1 + C_2) \epsilon^2
  - \frac{1}{2} (3 - 5 C_2) \epsilon \eta + \frac{1}{2} (3 - C_2)
  \xi,\\
  \alpha_{\rm S} & \simeq - \frac{1}{1 - \epsilon} \left( 2\xi + 8
    \epsilon^2 - 10 \epsilon \eta + \frac{7 C_2 - 9}{2} \, \epsilon
    \xi + \frac{3 - C_2}{2} \, \eta \xi \right),\\
    n_{\rm T} & \simeq -2 \epsilon - (3 + C_2) \epsilon^2 + (1 + C_2)
    \epsilon \eta,
\end{eqnarray}
where $C_2 = 4(\ln 2 + \gamma) - 5$.  The slow-roll parameters only
need to be evaluated at a field value $\phi_*$, corresponding to the
time when $k_*$ leaves the horizon.  $A_{\rm S}$ and $A_{\rm T}$ are
calculated from equations (\ref{eq:ps}) and (\ref{eq:pt}), and the
spectra follow directly.  One does not need to solve any differential
equations here, so the numerical implementation of this method is by
far the simplest of the three. However, due to the additional
assumption on the shape of the spectrum, it becomes increasingly
inaccurate the further one goes away from the pivot scale.

Apart from that, in the spirit of choosing a minimal prior one would
typically rule out only those models that break the $\epsilon < 1$
condition at the pivot scale, thus allowing regions in parameter space in
which inflation would break down even within the observable range and
making the prediction of the spectra for these models extremely
unreliable.

\section{Constraints from current data \label{sec:constraints}}

In this section we present the constraints on inflationary parameter
space from a selection of current observations, comprising CMB data
from the WMAP \cite{Hinshaw:2006ia,Page:2006hz}, Boomerang
\cite{Montroy:2005yx,Jones:2005yb,Piacentini:2005yq} and ACBAR
\cite{Reichardt:2008ay} experiments, complemented by the galaxy power
spectrum constructed from the luminous red galaxy sample of the Sloan
Digital Sky Survey \cite{Tegmark:2006az}. We analytically marginalise
over the luminous to dark matter bias $b^2$ and the nonlinear correction
parameter $Q_{\rm nl}$.

We consider a $\Lambda$CDM-model with eight free parameters, on which
we impose flat priors. Four of these parameters determine the initial
perturbation spectra: the scalar normalisation $\ln \left[ 10^{10}
A_{\rm S} \right]$, and the first three slow-roll parameters:
$\epsilon$, $\eta$ and $\xi$, evaluated at the pivot scale
$k_*=0.01~$Mpc$^{-1}$. We emphasise once more that the numerical
computation of perturbations does not refer to any slow-roll
expansion, and remains self-consistent even when the field is not
rolling very slowly. The fact of varying parameters which coincide
with the usual slow-roll parameters is just a choice of prior in
parameter space, which is particularly convenient for two reasons:
first, the posterior is well-behaved with respect to these parameters
and the convergence of the chains is achieved in a reasonable amount
of time; second, it facilitates comparison with other works.  The
remaining four parameters are the baryon density $\omega_b$, the cold
dark matter density $\omega_{\rm dm}$, the ratio of sound horizon to
angular diameter distance at decoupling $\theta_s$, and the optical
depth to reionisation $\tau$.  We use a modified version of the
Markov-Chain-Monte-Carlo code
\texttt{CosmoMC}~\cite{Lewis:1999bs,Lewis:2002ah} to infer constraints
on the free parameters of the model. The inflation module was made
publicly available by the authors of Ref.~\cite{Lesgourgues:2007aa} at
\url{http://wwwlapp.in2p3.fr/~valkenbu/inflationH/}.

\subsection{Expansion in $H$ vs.~expansion in $H^2$}

We first check the impact of changing the parameterisation of $H(\phi)$
(i.e., the precise ensemble of inflationary models considered) from a
Taylor-expansion of order 3 in $H(\phi)$ to the same expansion in
$H^2(\phi)$. In both cases, we used the same priors on inflationary
parameters: hence the difference only resides in the fact that
slightly different background evolutions can be achieved in both cases.
The differences are summarised in table~\ref{table:hvsh2} and turn out
to be very minor. This preliminary analysis shows that the parametric
form assumed for $H(\phi)$ within the observable window has a minor
impact. Significant differences could only be expected if the choice
of parameterisation of $H(\phi)$ would allow much more freedom in one
case than in the other.  

\begin{table}[t]
  \caption{\label{table:hvsh2}Minimal 95\%-credible intervals for the
    slow-roll parameters in the $H$- and $H^2$-expansion schemes,
    using the exact method for calculating the spectra.}\vskip5mm
    \hskip45mm \footnotesize{
\begin{tabular}{ccc}
  \br & $H$ & $H^2$ \\ \mr 
  $\epsilon$ & $\hphantom{-0.000}$0 $\to$ 0.028$\hphantom{000}$ & $\hphantom{-0.000}$0  $\to$  0.023 $\hphantom{000}$\\ 
  $\eta$ & -0.035 $\to$ 0.046  $\hphantom{.}$ & -0.035  $\to$ 0.039  $\hphantom{0}$\\
  $\xi$ & -0.0026 $\to$ 0.028$\hphantom{00}$ & -0.0053  $\to$  0.027  $\hphantom{00}$\\
 \br
\end{tabular}}
\end{table}


In the remaining part of the paper, we shall therefore stick to the
Taylor-expansion in $H(\phi)$ and perform three independent analyses,
calculating the primordial spectrum either by exactly solving the mode
equations, or using one of the two approximations discussed in
sections~\ref{subsec:apprI} and \ref{subsec:apprII}.

\subsection{Approximations vs.~exact spectra}

\begin{figure}
\includegraphics[height=\textwidth, angle=270]{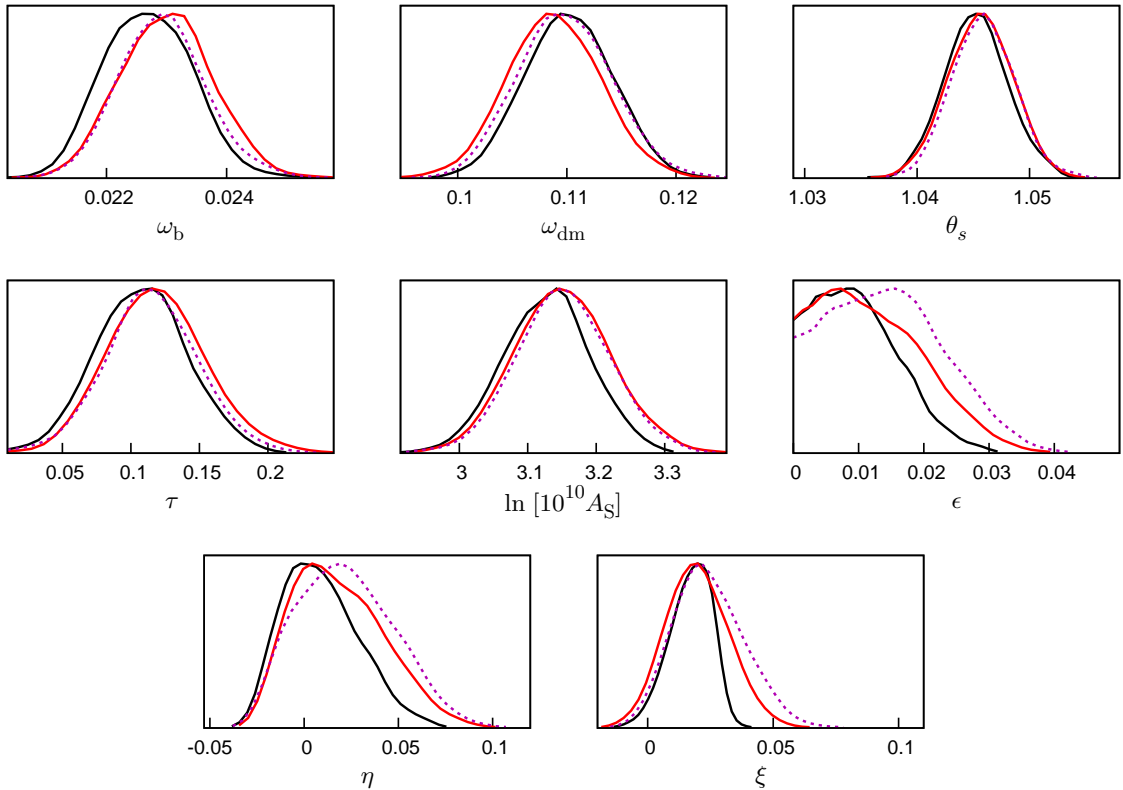}
\caption{This plot shows the one-dimensional marginalised posterior
  distributions for the free parameters of the model. The black lines
  represent the results of the exact solution of the mode equation,
  red lines are approximation~{\it I} and purple (dashed) lines
  correspond to approximation~{\it II}. \label{fig:hsrp}}
\end{figure}

\begin{figure}
\includegraphics[height=\textwidth, angle=270]{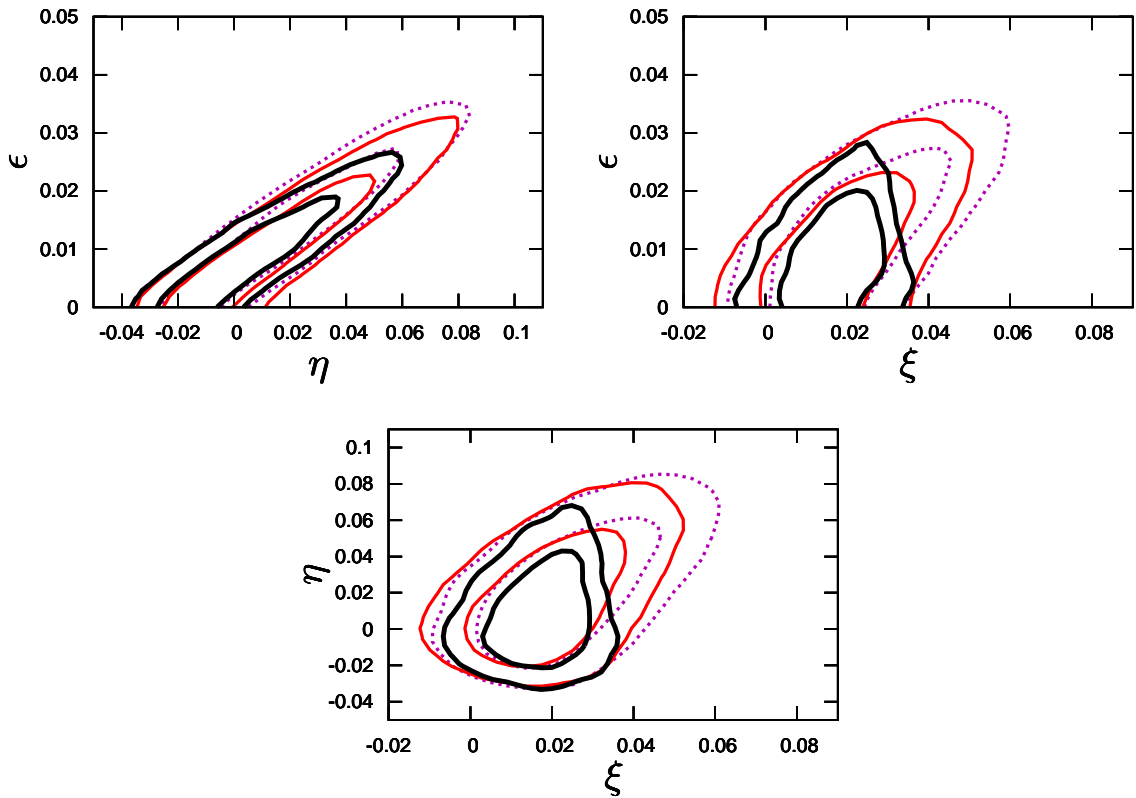}
\caption{This plot shows the 68\%- and 95\% credible regions
  of the two-dimensional marginalised posterior in the ($\epsilon,
  \eta$)- (top left), ($\epsilon,\xi$)- (top right), and
  ($\eta,\xi$)-planes (bottom). The black lines denote the results
  of the exact solution of the mode equation, red lines are
  approximation~{\it I} and thin lines correspond to approximation
  II. \label{fig:hsrp2d}}
\end{figure}

Our results are presented in figures~\ref{fig:hsrp} and
\ref{fig:hsrp2d}.  We do not find any significant differences in the
posterior probabilities of $\tau$, $\theta_s$, $\omega_{\rm b}$ and
$\omega_{\rm dm}$. The four parameters that determine the primordial
spectra, however, are more sensitive to the method used. Note that the
exact method produces tighter bounds on the slow-roll parameters,
particularly on $\xi$.

This also has important consequences on the inferred values of derived
phenomenological parameters, such as the spectral index and its
running. As can be seen from table~\ref{table:running}, the exact
method yields significantly stronger constraints on these two
parameters. 

\begin{table}[h]
  \caption{\label{table:running}Minimal 95\%-credible intervals for
    the spectral index and the running at a scale of $k_* = 0.01\ {\rm
      Mpc}^{-1}$. Note that these are derived parameters and the results
    are not independent of the choice of pivot scale.}\vskip5mm
\hskip25mm \footnotesize{
\begin{tabular}{cccc}
  \br
  & exact & approximation~{\it I} & approximation~{\it II} \\
  \mr
  $n_{\rm S}$ & $0.959 \to 1.049$ & $0.960 \to 1.078$ & $0.960 \to 1.087$\\
  $\alpha_{\rm S}$ & $-0.063 \to 0.001\hphantom{-}$ & $-0.084 \to
  0.009\hphantom{-}$ & $-0.098 \to 0.003\hphantom{-}$\\ 
  \br
\end{tabular}}
\end{table}

\section{\label{sec:diff}Why the difference?}

There are potentially two reasons for these observed discrepancies.
The first one is that the accuracy of the approximations could be
insufficient within their respective ``allowed'' parameter space and
lead to a serious bias in the parameter estimates. Note that the
discrepancy occurs mostly in regions of parameter space where $\xi$ is
large. The larger $\xi$, the more one would expect the accuracy of the
approximations to degrade. However, given that the approximations are
expected to be accurate to order $\xi$, i.e., not worse than $\sim \! 
10\%$, an effect as large as the one we observe seems rather unlikely.

The second reason is slightly more subtle: as we discussed in section
\ref{sec:perturbations}, the three methods differ in their {\it
implicit} prior on the space of models. While approximation~{\it II}
requires $\epsilon < 1$ only at the pivot scale, approximation~{\it I}
needs us to demand that this condition be fulfilled in the entire
observable window of $\sim \! 10$ $e$-folds, corresponding to Hubble
exit for modes in the $[k_{\rm min}, k_{\rm max}]$ range. In the exact
numerical approach, we require $\epsilon < 1$ for the whole
integration time, which starts when $aH/k_{\rm min}=1/50$ instead of
one, and ends when $|\xi_{\rm S,T}|/z_{\rm S,T}$ freezes out, i.e., a
few $e$-foldings after $aH/k = 1$, corresponding to an even more
restrictive prior.

It was pointed out in Refs.~\cite{Easther:2006tv,Malquarti:2003ia}
that for models with large positive values of $\xi$ ($> 0.05$) and no
higher derivatives, inflation tends to end within a few $e$-foldings
of the pivot scale leaving the horizon\footnote{If higher derivatives
are present this conclusion can be weakened, see, e.g.,
\cite{Makarov:2005uh,Ballesteros:2005eg}.}. This is consistent with
our results, since the more restrictive priors lead to tighter bounds
on~$\xi$.

\subsection{The prior issue}

To verify that the differences actually stem from the choice of priors
and not from a lack of accuracy, we post-processed our Markov chains
of the approximate methods, discarding all models for which inflation
is interrupted in the range of wavelengths required for the exact
calculation.

In a first step, we remove only those models for which the
inflationary condition is violated {\it before} the pivot scale leaves
the horizon, when $aH$ is in the range $[k_{\rm min}/50,k_*]$. These
are models for which the assumption of the Bunch-Davies vacuum initial
condition is violated at least for the largest observable wavelengths.
Only a mere 0.02\% of the models in the chains using
approximation~{\it I}, and 0.01\% for approximation~{\it II}, fall
victim to the cut\footnote{Here, and in the following, we quote a
weighted fraction of models, i.e., $(\sum_j w_j^{\rm bad})/\sum_i
w_i)$, where $w_i$ are the statistical weights of the points in the
Markov chains, and $w_j^{\rm bad}$ are the weights of the models
killed by the prior.}. This is probably connected to the dislike of the
data for models with large negative $\xi$, which is required if we
want inflation to start only just before the observable range.


Imposing the same additional prior as in the exact method (that
inflation holds till the time of freeze out for each mode), $\sim
20\%$ of the approximation~{\it I} points and $\sim 34\%$ of the
approximation~{\it II} points are removed. After weeding out the bad
models, the bounds of the approximations perfectly agree with the ones
derived using the exact method, their marginalised posteriors are
virtually indistinguishable. This confirms our suspicion that the
different priors are responsible for the discrepancy between the
methods.

\subsection{Comparison of accuracy}

Having seen that the prior plays a very important role, it is
nonetheless interesting to take a closer look at how the
approximations compare to the exact method in terms of accuracy.

In order to compare the accuracy of the three methods described
previously, we took the 95\% best-fitting spectra obtained using
approximations {\it I} and {\it II} and for each model in the chains
we again computed the curvature spectrum in either of the
approximations and numerically in order to compare. We searched for
the maximum discrepancy between the approximated spectrum (with method
{\it I} or {\it II}) and the numerical one, in each of the two ranges
$[ k_{\rm min}, k_* ]$ and $[k_*, k_{\rm max}]$, with $k_{\rm min} =
3 \times 10^{-6}$, $k_* = 0.01$, $k_{\rm max} = 1.2$, in units of 1/Mpc,
corresponding to the range of wavelengths the data are most sensitive
to. Note that the spectrum for a model can only be computed
numerically if the model meets the prior condition on the duration of
inflation. Hence the comparison done here is for models that are
already preselected by that particular prior, whereas in the actual
chains many points exist that give a much larger discrepancy due to
the different prior.

\begin{figure}
\includegraphics[width=0.45\textwidth]{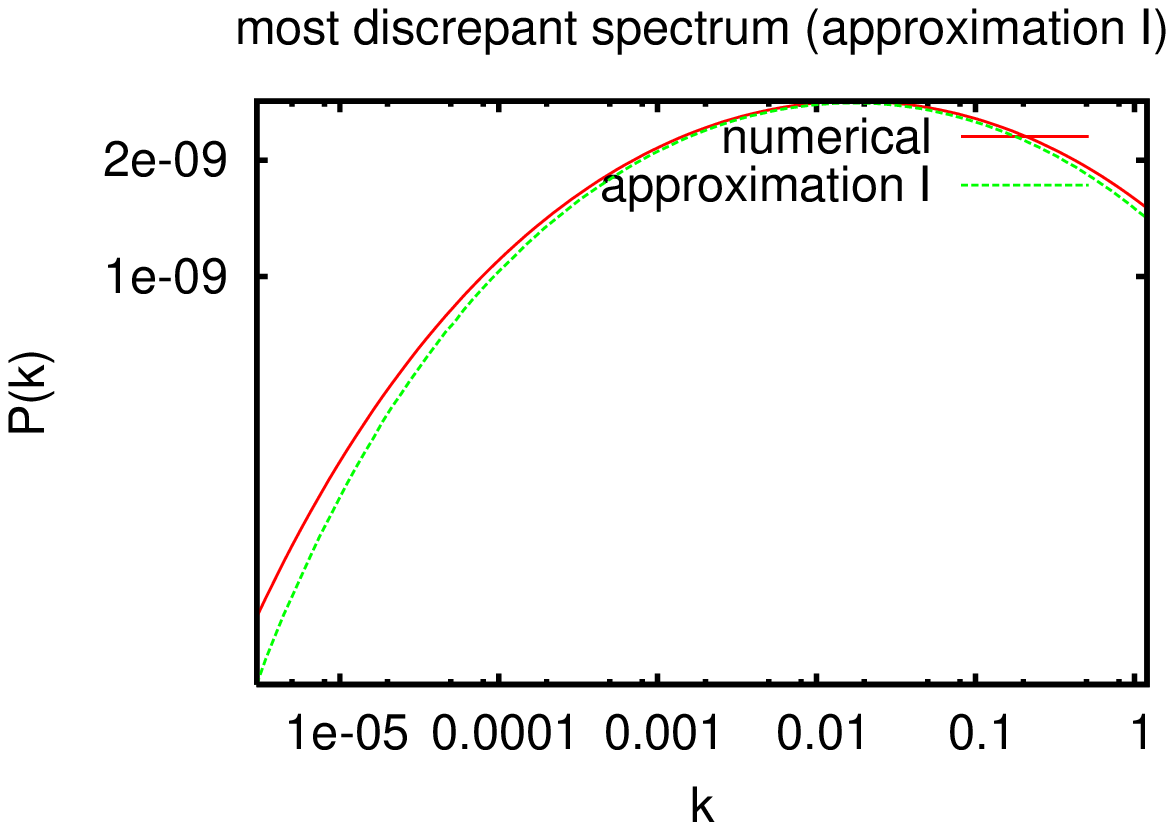}
\includegraphics[width=0.45\textwidth]{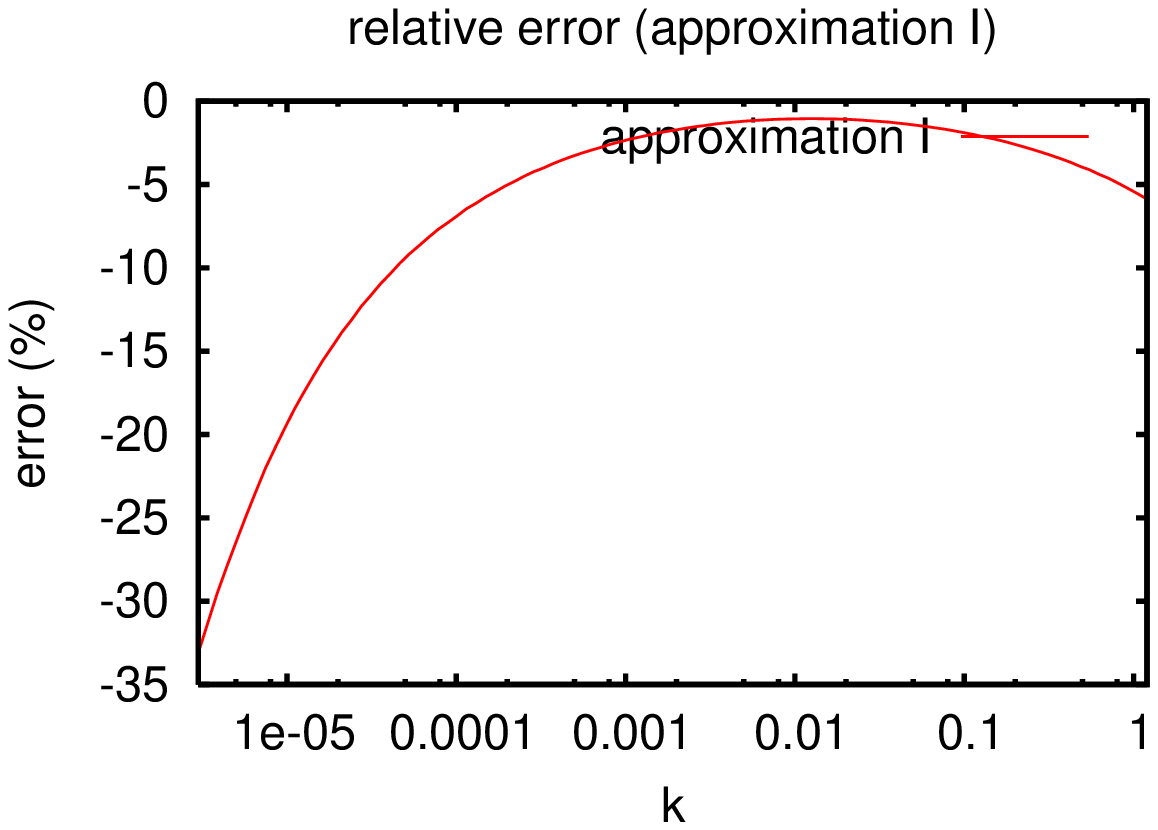}\\
\includegraphics[width=0.45\textwidth]{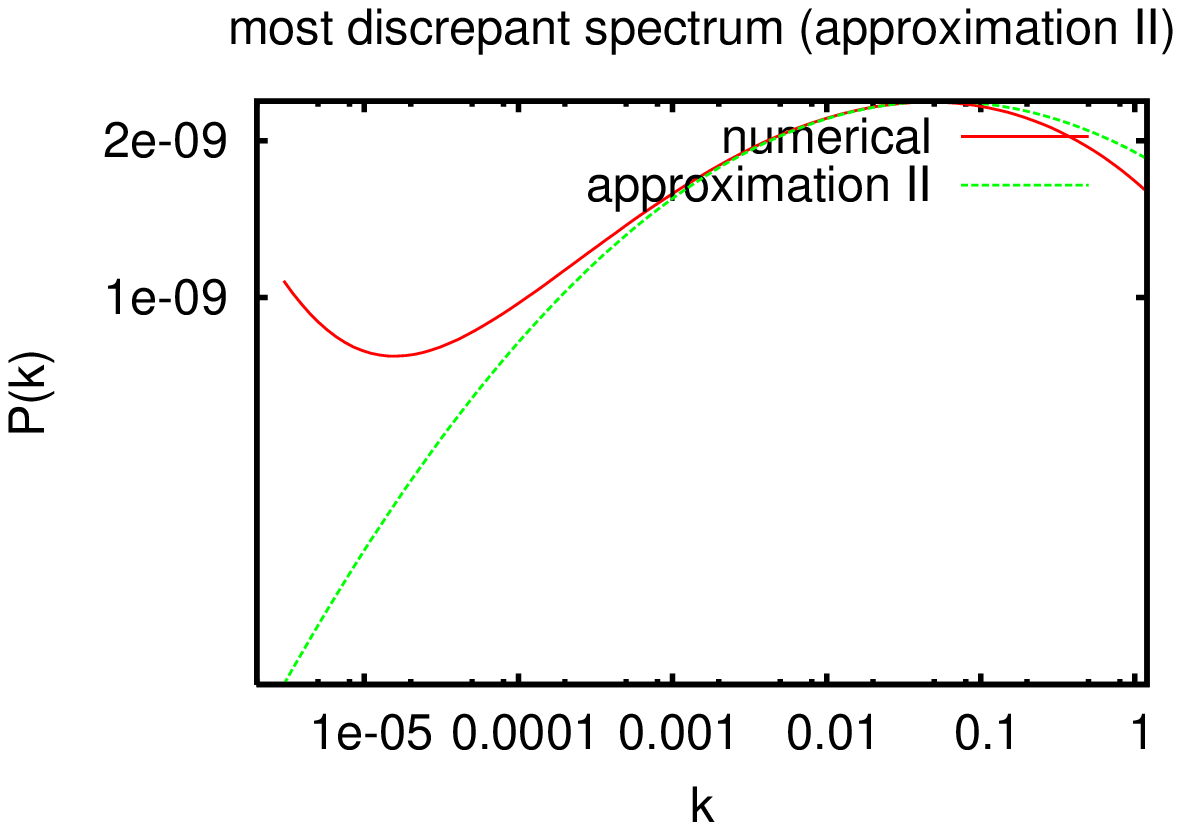}
\includegraphics[width=0.45\textwidth]{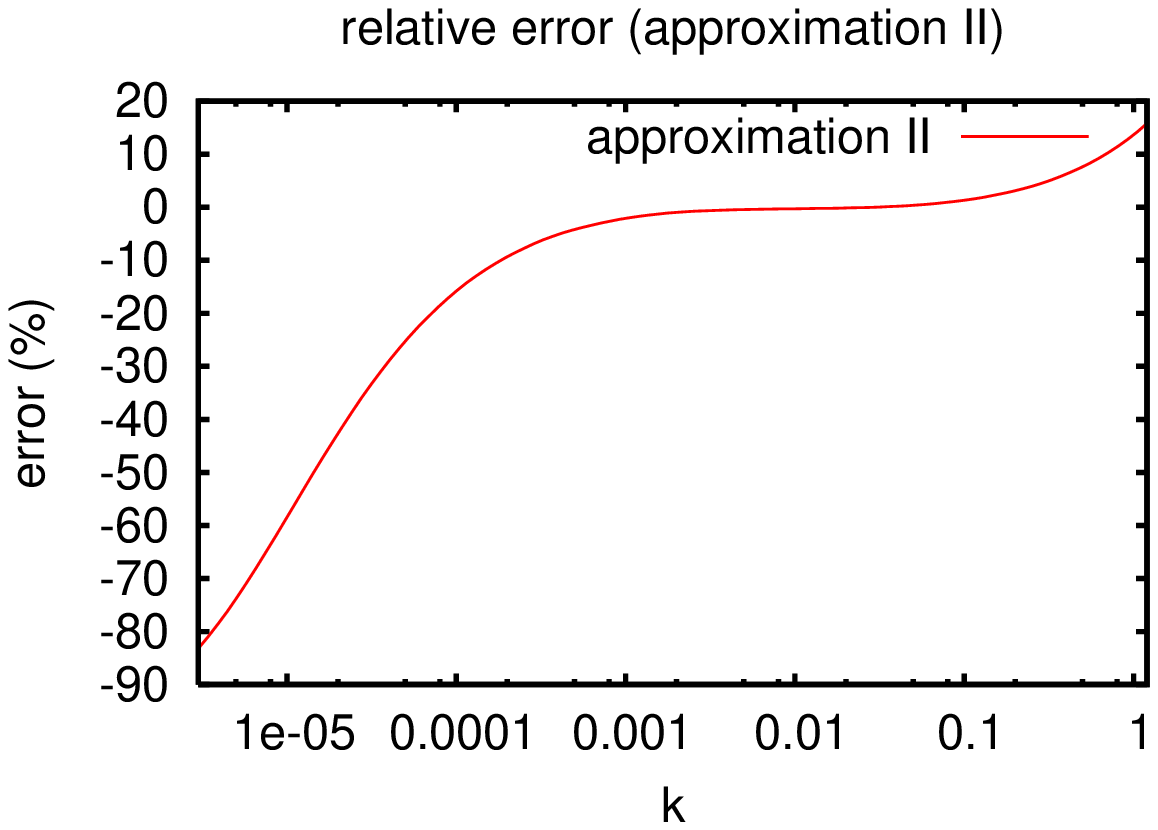}
\caption{{\it Left:} Curvature spectrum obtained from the exact
 numerical method or from approximation {\it I} or {\it II}, for the
 most discrepant models in the range \mbox{$k \in [3\times 10^{-6},
 1.2]$~Mpc$^{-1}$} selected among the 95\% best-fitting spectra. {\it
 Right:} The corresponding relative differences. The discrepancy is
 large only for scales close to $k_{\rm min}$, which have a relatively
 small statistical weight.
\label{ill}}
\end{figure}

\begin{table}
\caption{Value of the slow-roll parameters at the pivot scale
for the models referred to as A and B in the text. The corresponding curvature
spectra computed with each method are shown in  Figure~\ref{ill}.
\label{abcd}}\vskip5mm
\hskip30mm \footnotesize{
\begin{tabular}{cccc}
model & $\epsilon$ & $\eta$ & $\xi$ \\ \hline
A & $1.1446\times 10^{-3}$ & $-4.1728\times 10^{-3}$ & $3.4911\times 10^{-2}$ \\
B & $2.6159\times 10^{-2}$ & $7.5254\times 10^{-2}$ & $3.3839\times 10^{-2}$ 
\end{tabular}}
\end{table}

In Figure~\ref{ill} we plot the most discrepant models 
in the $[k_*, k_{\rm max}]$ range, that we call A
(for approximation~{\it I}) and B (for approximation~{\it II}).  The
corresponding slow-roll parameters evaluated at the pivot scale are
given in Table~\ref{abcd}.  For approximation~{\it II} we find a
maximum discrepancy in ${\cal P}_{\cal R}$ of 83\% below $k_*$, and
19\% above (model B).  For {\it I}, the difference reduces to 33\%
below $k_*$ and 8\% above (model A).  So, approximation~{\it I} is
doing better on both sides of the pivot scale.

Indeed, it appears that approximation~{\it I} captures the spectrum
shape very well, but not its amplitude. This is not surprising since
in {\it I}, the expression for the amplitude is first-order in the
slow-roll expansion. The next-order contribution to this approximation
should include the parameter $\xi$, with a positive coefficient. Since
the data allows large values of $|\xi|$ only when $\xi$ is positive,
approximation~{\it I} yields a systematic underestimation of the
amplitude. This approach could be straightforwardly improved by
computing the curvature amplitude at the next order.  In contrast,
approximation~{\it II} tends to give the wrong shape, but since it
includes one more order in the slow-roll expansion it can give a
better estimate of the amplitude for $k$ not too far from $k_*$.

However, the main result of this section is that the difference
between the various spectra is very small, since a large discrepancy
is only encountered on scales close to $k_{\rm min}$, which have a
relatively small statistical weight in the process of accepting or
rejecting a model. For both approximations the error in the larger
part of the spectrum is of the order of 5\%.  Current data does not
reach such good sensitivity, especially if we keep in mind that for
the largest $k$ values the curvature spectrum is mainly constrained by
the SDSS data, which is always marginalised over an unknown bias
parameter. For the particular models shown here, the difference in the
effective $\chi^2$ obtained when fitting either the approximated or
the numerical spectrum to the data is $|\Delta\chi_{\rm eff}^2|=6.7$
for approximation~{\it I} (model A) and $|\Delta\chi_{\rm eff}^2|=5.9$
for approximation~{\it II} (model B). One should keep in mind though,
that these are just the most extreme deviations, at the edge of
allowed parameter space. On average, the inaccuracies are too small to
have a significant effect on the inferred bounds.
However, with future datasets one can expect the $|\Delta\chi_{\rm
eff}^2|$ to become even larger, possibly resulting in biased
estimates. Hence, we recommend using the exact numerical approach,
since it does not make the analysis longer or more difficult.

The conclusions reached in this section apply to a particular class of
inflationary models, namely those described by Eq.~(\ref{taylor}) with
parameter ranges limited by current WMAP and SDSS results. Allowing
for more freedom in $H(\phi)$, one would expect stronger deviations
between the analytical and numerical approaches. Conversely, imposing
a constraint on the total number of $e$-folds for relevant inflation,
one would select models which are deeper within slow-roll and obtain
even smaller discrepancies. We have limited our discussion to the
scalar spectrum, since there is, at present, no evidence for anything
but a subdominant tensor contribution in the available data. This may
of course change if a primordial $B$-mode polarisation of the CMB is
detected in the future.

\section{Discussion}\label{sec:conclusions}

We have compared various alternative methods for putting constraints
on the observable window of inflationary dynamics, assuming
single-field inflation with a smooth behaviour. One could fear that
the results would depend very much on the way to compute the spectrum,
or on the parameterisation of $H(\phi)$ (i.e., the ensemble of models
considered). We point out that with current data these differences are
subdominant. The results are mainly affected by the exact prior on the
background. By focusing on the allowed window, one hopes to be
conservative and to get results dictated by the data only; however
these results are very sensitive to the edges of the interval in field
space over which accelerated expansion is required.  In other words,
the upper bound on $\xi$ is given by this constraint rather than the
shape of the perturbation spectrum. Current analyses agree with each
other, but only within a factor two, due to this difference.  This
dependence of the final results on the choice of prior may sound
worrisome, particularly if it is one's aim to keep the analysis as
general as possible.

It seems as though the priors we initially chose to use with
approximations~{\it I} and {\it II} are less restrictive, and would
therefore lead to a more conservative result. Unfortunately however,
combining these methods with their respective priors on the space of
allowed models is fraught with a severe consistency problem. Both
priors allow models in regions of parameter space where the
approximations are known to break down, and yield results for the
spectrum that cannot be trusted. Approximation~{\it II}, for instance,
allows models in which inflation is interrupted within the observable
range.  Such an event would lead to very distinct signatures, like a
cutoff, yet the approximation would still predict a smooth spectrum.
It is therefore sensible to expect that inflation lasted at least over
the whole observable range. But even that will not be sufficient: if
inflation started only just before the observable range, the
assumption that the modes start out in the Bunch-Davies vacuum can no
longer be justified, and the approximations fail. If, on the other
hand, inflation ends just after the smallest observable scale leaves
the horizon, the corresponding mode will not have time to freeze
out. In fact it would re-enter the horizon right away, and there is no
telling (without making further assumptions) in what shape the
spectrum would arrive at later times when it is relevant for the
determination of the CMB anisotropy spectra.

Hence, it is reasonable to demand a proper vacuum initial condition
and a freeze-out of the modes. One could in principle further limit
the space of allowed models by constraining the minimum number of
$e$-folds before the end of inflation to a certain number, usually
taken to be $\geq 30$. However, this would require a daring
extrapolation of our simple Taylor-expansion over a huge range of
$e$-folds, where even a tiny higher derivative of the Hubble parameter
would eventually take over. In this paper, therefore, we did not want
to make any additional assumptions about what happens after the
freeze-out. 

There are, however, two points at which the prior of the
self-consistent numerical approach is slightly arbitrary,
corresponding to the two end points of the interval over which we
track the background dynamics. The first one is the large scale end,
determined by the time at which we choose the initial conditions for
the $k_{\rm min}$. We picked $k_{\rm min}/50$ as a starting point, but
other choices may have been equally good.  Fortunately, the data
conspire to make this choice have little impact on the final results:
only about 0.1\% of the models in our chains generated with the laxer
prior can be rejected due to inflation starting ``too late''.

The more critical issue is the small scale end of the interval. Its
choice is connected with the question when a mode can be considered to
have frozen out, which is set by the limiting value of $[ {\rm d} \ln
{\cal P}_{{\cal R}, T} / {\rm d} \ln a ]$ at which we stop
integrating. The final results for the posterior are mildly dependent
on the choice of this limit, which should eventually be chosen such
that the resultant uncertainty in the spectra is smaller than the
sensitivity of the data. For this reason, we choose here $[ {\rm d}
\ln {\cal P}_{{\cal R}, T} / {\rm d} \ln a ] < 10^{-3}$, corresponding
to a 0.1\% accuracy in the power spectra.

We would like to emphasize once again that the differences in the
results are not inherent to the approximations used, but rather due to
the attempt to implement them with a minimal prior. Our results also
show that if one were to impose a non-minimal prior on the number of
$e$-foldings beyond the observable range, the approximations would
lead to the same results as the exact method.

In the future, we expect more robust constraints from high-precision
experiments, such as, e.g., the Planck satellite. In turn the
difference between the various methods for computing the spectra will
become even more relevant. In the light of our results, we recommend
using the exact numerical approach for a self-consistent analysis of
inflationary dynamics.

\section*{Acknowledgments}

We wish to thank P.~Adshead, R.~Easther and H.~Peiris for very
stimulating discussions, and A.~Starobinsky for enlightening comments.
JH is supported by the ANR (Agence Nationale de la Recherche). WV is
supported by the EU 6th Framework Marie Curie Research and Training
network ``UniverseNet'' (MRTN-CT-2006-035863). Numerical simulations
were performed on the MUST cluster at LAPP (CNRS \& Universit\'e de
Savoie).

\section*{References}
\bibliographystyle{iopart-num}
\bibliography{refs}

\end{document}